\documentclass{article}
\usepackage{graphicx} 
\usepackage{authblk}
\usepackage[utf8]{inputenc}
\usepackage{cite}
\usepackage{subcaption}
\usepackage{amsmath}
\usepackage{breqn}
\usepackage[a4paper,margin=0.5in]{geometry}

\title{Nonlinear viscous gravity-capillary surface waves at arbitrary wavelengths}

\author[1]{Arash Ghahraman}
\author[1]{Gyula Bene}

\affil{{ghahraman.arash@ttk.elte.hu}, {bene.gyula@ttk.elte.hu}}
\affil[1]{Department of Theoretical Physics, Eötvös University, Pázmány Péter sétány 1/A, 1117 Budapest, Hungary}

\date{March 26, 2023}

\begin{document}

\maketitle

\abstract{
This paper presents the second-order perturbation theory of the Navier-Stokes equations for free surface flows, with the wave amplitude considered as the perturbation parameter. Gravity-capillary surface waves in incompressible viscous fluids are subjected. The results provide a systematic derivation of a nonlinear surface wave equation that fully takes into account dispersion, while nonlinearity is included in the leading order. However, the presence of infinitely many overdamped modes has been neglected, and only the two least damped modes are considered \cite{ghahraman2023bifurcation}. Finally, to describe the elevations and evolution of the surface wave, we introduce a differential equation.

}

\section{Introduction}


Studying free surface flows has a long history.
Various equations have been proposed to model propagation of surface waves in the presence or absence of viscosity \cite{ref43_lamb1924hydrodynamics, ref12_boussinesq1895lois, segur2005stabilizing, joseph2004dissipation,wang2006purely}. The goal is to find reduced models, particularly in size on simplified domains with as little number of fields as possible, should they be valid only in an asymptotic regime \cite{ref01_Meur2015derivation}.  Boussinesq \cite{ref12_boussinesq1895lois} and Lamb \cite{ref43_lamb1924hydrodynamics} studied the effect of viscosity on free surface waves. They concentrated on linearized NS equations on deep-water and computed the dispersion relation. Basset \cite{ref13_basset1888treatise} also worked on viscous damping of water waves. The first formal dissipative KdV equation  was presented by Ott and Sudan \cite{ref14_ott1970damping}. In 1974, an equation was proposed by Whitham as an alternative to the well-known KdV equation \cite{whitham1974linear}. Whitham proposed to use the same nonlinearity as the KdV equation, but coupled with a linear term which mimics the linear dispersion relation of the full water-wave problem. 
Whitham proposed a general approach that relies on the consideration of two scales of time: the fast oscillations of the wave and the slow changes of its parameters \cite{whitham1974linear, kamchatnov2004whitham}. The equations describing the slow evolution of the wave parameters can be obtained by averaging over the fast oscillations, leading to a set of first-order partial differential equations known as the Whitham equations. However, it is important to note that small perturbations, which are ignored in completely integrable approximations, can significantly affect the long-term behavior of the system. Hence, there is a need to develop a perturbation approach to the higher orders \cite{li2020solitary, demiray2017exact, wang2022solitary}.


Dias et al. 
\cite{ref22_dias2008theory} utilized the Navier-Stokes equations and incompressible potential flow theory to derive a set of equations that describe weakly damped free surface waves (DDZ). The dynamic and kinematic boundary conditions were augmented with viscous terms, and Dutykh and Dias \cite{ref23_dutykh2007viscous, ref22_dias2008theory} demonstrated that only in this situation could the phase velocity remain unaltered by the inclusion of dissipation. The DDZ system has proven useful for numerically simulating water waves with viscosity \cite{carter2019comparison, granero2019models, mouassom2021effects,armaroli2018viscous} and for establishing a well-posedness theory \cite{ambrose2012well,ngom2018well}.

The aim of this study is to construct a nonlinear viscous model which is generalized by considering surface tension and nonlinear terms in Navier stokes equation and boundary conditions that can accurately simulate the evolution of surface waves. In the present paper we choose an approach that allows both taking into account dispersion fully (at arbitrary depth) and a systematic calculation of nonlinear terms. To this end we solve Navier-Stokes equations for the surface wave problem perturbatively to second order, with wave amplitude being a small parameter. The result allows us to derive a self-consistent nonlinear integro-differential equation for surface elevations. The first order (linear) approximation has already been extensively studied in \cite{ghahraman2023bifurcation}. It was found that at any given horizontal wave number infinitely many modes exists, but only those with the smallest damping rate may describe propagating modes. In a thin enough fluid layer no propagation occurs at all.
Even in any thicker fluid layers, propagation of very short and very long waves is forbidden. Keeping the least damped two modes, the following bidirectional evolution equation has been introduced in wave number space, valid independently whether propagation is possible or not:
 
\begin{equation}
    \frac{\partial^2 \zeta}{\partial t^2}+i (\Omega^{\left ( 1 \right )}\left ( k \right )+\Omega^{\left ( 2 \right )}\left ( k \right )) \frac{\partial \zeta}{\partial t}-\Omega^{\left ( 1 \right )}\left ( k \right ) \Omega^{\left ( 2 \right )}\left ( k \right ) \zeta =0\;.
    \label{linear_eq}
\end{equation}

Here $\Omega^{(1)}$ and $\Omega^{(2)}$ are angular frequencies of the two modes. 

The plan of the paper is as follows. In Section 2 a formulation of the problem is given. First order and second order approximations are discussed in Sections 3 and 4, respectively. In Section 5 we derive the surface wave equation based on perturbation theory. The results are summarized and discussed in the concluding Section 6. Some longer expressions are relegated to the Appendix. Note that there are three further expressions whose explicit length is over 100 pages, these are made available as supplementary material. 

\section{Formulation of the problem}
Consider the irrotational flow of an incompressible inviscid fluid with a free surface. A Cartesian coordinate system is adopted, with the x-axis located on the still-water plane and with the z-axis pointing vertically upwards. The fluid domain is bounded by the bed at $z=-h$ and the free surface at $z=\zeta(x,t)$.
The Navier-Stokes (NS) equations in the bulk are:

\begin{equation}
    \frac{\partial \boldsymbol{V}}{\partial t}+(\boldsymbol{V} \cdot \nabla )\boldsymbol{V}-\nu \triangle \boldsymbol{V}=\nabla (-\frac{p}{\rho}-gz) \label{NS_eq}
\end{equation}

Since the right hand side is a full gradient, we have

\begin{equation}
\frac{\partial }{\partial z} \left(\frac{\partial u}{\partial t}+u \frac{
  \partial u}{\partial x}+w\frac{
  \partial u}{\partial z}-\nu
\triangle  { u} \right)
=\frac{\partial }{\partial x}\left(\frac{\partial { w}}{\partial t}+u\frac{
  \partial w}{\partial x}+w\frac{
  \partial w}{\partial z}-\nu
\triangle  { w} \right) \label{NS2}
\end{equation}

with bottom boundary conditions

\begin{equation}
u(z=-h)=w(z=-h)=0\ \label{btm_bc}
\end{equation}

and horizontal and vertical components of net force at the surface (up to second order precision in amplitudes):
\begin{dmath}
\rho \nu\left(\frac{\partial u}{\partial z}+\frac{\partial w}{\partial
  x}\right)=\frac{\partial \zeta}{\partial x}\left(p_0-p+2\rho \nu
\frac{\partial u}{\partial x}-\sigma
\frac{\partial^2\zeta}{\partial x^2}\right)-\zeta \frac{\partial}{\partial z}\left[\rho \nu\left(\frac{\partial u}{\partial z}+\frac{\partial w}{\partial x}\right)-\frac{\partial \zeta}{\partial x}(p_0-p)\right]\;,
\label{sbc_h}
\end{dmath}

\begin{equation}
    p_0-p+2\rho \nu \frac{\partial w}{\partial z}-\sigma
\frac{\partial^2\zeta}{\partial x^2}=-\zeta \frac{\partial}{\partial z}\left(p_0-p+2\rho \nu \frac{\partial w}{\partial z}\right)+\frac{\partial \zeta}{\partial x}\rho \nu\left(\frac{\partial u}{\partial z}+\frac{\partial w}{\partial
  x}\right)\;.
\label{sbc_v}
\end{equation}

Further, we have the kinematic condition for the surface
\begin{equation}
    \frac{\partial \zeta}{\partial t}-w=-u\frac{\partial \zeta}{\partial x}+\zeta \frac{\partial w}{\partial z}\;.
\label{sbc_kinbc}
\end{equation}

By eliminating the pressure from Eqs. (\ref{sbc_h}) and (\ref{sbc_v}) with the help of the Navier-Stokes equations, we have the following equations respectively.

\begin{dmath}
 \left(\frac{\partial \zeta}{\partial x}\right)^{2}\left(\frac{\partial { u}}{\partial t}-\nu
\triangle  { u} +2\nu
\frac{\partial^2 u}{\partial x^2}-\frac{\sigma}{\rho}
\frac{\partial^3\zeta}{\partial x^3}\right) =  -\frac{\partial^2 \zeta}{\partial x^2}
\left\{\nu\left[\frac{\partial u}{\partial z}+\frac{\partial w}{\partial
  x}+\zeta\left(\frac{\partial^2 u}{\partial z^2}+\frac{\partial^2 w}{\partial
    x\partial z}\right)\right]- g \zeta\frac{\partial \zeta}{\partial
  x}\right\}\nonumber\\
 + \left(\frac{\partial \zeta}{\partial x}\right)\left\{
\nu\left[\frac{\partial^2 u}{\partial x \partial z}+\frac{\partial^2 w}{\partial
  x^2}+\frac{\partial \zeta}{\partial
  x}\left(\frac{\partial^2 u}{\partial z^2}+\frac{\partial^2 w}{\partial
    x\partial z}\right)+\zeta\left(\frac{\partial^3 u}{\partial x\partial z^2}+\frac{\partial^3 w}{\partial
    x^2\partial z}\right)\right]- g \zeta\frac{\partial^2 \zeta}{\partial x^2}-g\left(\frac{\partial \zeta}{\partial
  x}\right)^2\right\}
\label{f16d}
\end{dmath}

\begin{dmath}
    \frac{\partial { u}}{\partial t}+u\frac{
  \partial u}{\partial x}+w\frac{
  \partial u}{\partial z}-\nu
\triangle  { u}+2 \nu \frac{\partial^2 w }{\partial x \partial z }-\frac{\sigma }{\rho }
\frac{\partial^3\zeta }{\partial x^3 }=-\zeta  \left(\frac{\partial^2 {
    w}}{\partial t \partial x}-\nu\frac{\partial^3 w}{\partial
  x^3}+\nu\frac{\partial^3 w}{\partial x \partial z^2}
\right)\nonumber\\
-\frac{\partial \zeta}{\partial x} \left(g+\frac{\partial { w}}{\partial t}
-2 \nu  \frac{\partial^2 w}{\partial x^2}+ \nu  \frac{\partial^2 w}{\partial z^2}- \nu  \frac{\partial^2 u}{\partial x\partial z}
\right)
+\nu\frac{\partial^2 \zeta}{\partial x^2} \left(\frac{\partial u}{\partial z}+\frac{\partial w}{\partial  x}\right)
\label{f17b}
\end{dmath}

As a result, the main equations in this study are Eqs. (\ref{NS2}), (\ref{btm_bc}), (\ref{sbc_kinbc}), (\ref{f16d}) and (\ref{f17b}). 

Assuming incmpressible fluid, we may express the velocity components in
terms of the
stream function as
\begin{eqnarray}
  u & = & -\frac{\partial \psi}{\partial z} \label{g1a} \\  
  w & = & \frac{\partial \psi}{\partial x}
\label{g1b}
\end{eqnarray}

Inserting Eqs. (\ref{g1a}) and (\ref{g1b}) into the previous equations, we get
\begin{eqnarray}
  \frac{\partial \triangle \psi}{\partial t}-\nu \triangle \triangle\psi
  =\frac{\partial \psi}{\partial x}\frac{\partial \triangle \psi}{\partial z}
  -\frac{\partial \psi}{\partial z}\frac{\partial \triangle \psi}{\partial x}
\label{g2}
\end{eqnarray}
in the bulk, and
\begin{eqnarray}
  \frac{\partial \psi}{\partial x} & = & 0 \label{g3a}\\
  \frac{\partial \psi}{\partial z} & = & 0 
\label{g3}
\end{eqnarray}
at the bottom (\(z=-h\)). The surface boundary conditions are (at \(z=0\)):
\begin{dmath}
\frac{\partial \zeta}{\partial t}-\frac{\partial \psi}{\partial x}=\frac{\partial \psi}{\partial z}\frac{\partial \zeta}{\partial x}+\zeta \frac{\partial^2 \psi}{\partial x\partial z}\;,
\label{g4}
\end{dmath}

\begin{dmath}
\nu\left[\frac{\partial^2 \zeta}{\partial x^2}\left(\frac{\partial^2
    \psi}{\partial x^2}-\frac{\partial^2 \psi}{\partial z^2}\right)
  -\frac{\partial \zeta}{\partial x}\left(\frac{\partial^3 \psi}{\partial x^3}-\frac{\partial^3 \psi}{\partial x\partial z^2}\right)\right]
=-g\left(\frac{\partial \zeta}{\partial x}\right)^3+\left(\frac{\partial
  \zeta}{\partial x}\right)^2\left[\frac{\partial^2 \psi}{\partial t\partial z}
+\nu\left(\frac{\partial^3 \psi}{\partial x^2\partial z}-\frac{\partial^3 \psi}{\partial z^3}\right)+\frac{\sigma}{\rho}
\frac{\partial^3\zeta}{\partial x^3}
\right]\nonumber\\
-\nu \zeta \frac{\partial^2 \zeta}{\partial x^2}\left(\frac{\partial^3
  \psi}{\partial x^2\partial z}-\frac{\partial^3 \psi}{\partial z^3}\right)
+\nu \zeta \frac{\partial \zeta}{\partial x}\left(\frac{\partial^4
  \psi}{\partial x^3\partial z}-\frac{\partial^4 \psi}{\partial x\partial z^3}\right)
\;,
\label{g5}
\end{dmath}

and
\begin{dmath}
  -\frac{\partial^2 \psi}{\partial t\partial z}+3\nu \frac{\partial^2
    \psi}{\partial x^2\partial z}
  +\nu \frac{\partial^2 \psi}{\partial z^3}+g\frac{\partial \zeta}{\partial x}-\frac{\sigma }{\rho }\frac{\partial^3 \zeta}{\partial x^3}
=\frac{\partial \psi}{\partial x}\frac{\partial^2 \psi}{\partial z^2}-\frac{\partial \psi}{\partial z}\frac{\partial^2 \psi}{\partial x\partial z}
\nonumber\\
+\zeta\left[-\frac{\partial^3 \psi}{\partial t\partial
    x^2}+\nu\left(\frac{\partial^4
  \psi}{\partial x^4}-\frac{\partial^4 \psi}{\partial x^2\partial z^2}\right)\right]
+\frac{\partial \zeta}{\partial x}\left[-\frac{\partial^2 \psi}{\partial t\partial
    x}+2\nu\left(\frac{\partial^3
  \psi}{\partial x^3}-\frac{\partial^3 \psi}{\partial x\partial z^2}\right)\right]
+\nu \frac{\partial^2 \zeta}{\partial x^2}\left(\frac{\partial^2
  \psi}{\partial x^2}-\frac{\partial^2 \psi}{\partial z^2}\right)
\;.
\label{g6}
\end{dmath}

We are looking for a solution in a form
\begin{eqnarray}
  \psi & = & \psi^{[1]}+\psi^{[2]}\;, \label{g700}\\
      \zeta & = & \zeta^{[1]}+\zeta^{[2]}\;,
\label{g7}
\end{eqnarray}

where the superscript refers to the order. 

\section{First Order Solutions}

Inserting decompositions (\ref{g700}), (\ref{g7}) into
the equations, we have in first order
\begin{eqnarray}
  \frac{\partial \triangle \psi^{[1]}}{\partial t}-\nu \triangle \triangle\psi^{[1]}
  =0
\label{g2e1}
\end{eqnarray}
in the bulk (\(-h\le z \le 0\)), 
\begin{eqnarray}
  \frac{\partial \psi^{[1]}}{\partial x} & = & 0 \label{g3ae1}\\
  \frac{\partial \psi^{[1]}}{\partial z} & = & 0 
\label{g3e1}
\end{eqnarray}
at the bottom (\(z=-h\)), and 
\begin{eqnarray}
\frac{\partial \zeta^{[1]}}{\partial t}-\frac{\partial \psi^{[1]}}{\partial x}=0\;,
\label{g4e1}
\end{eqnarray}
\begin{eqnarray}
\frac{\partial^2
    \psi^{[1]}}{\partial x^2}-\frac{\partial^2 \psi^{[1]}}{\partial z^2}=0
\;,
\label{g5e1}
\end{eqnarray}
\begin{eqnarray}
  -\frac{\partial^2 \psi^{[1]}}{\partial t\partial z}+3\nu \frac{\partial^2
    \psi^{[1]}}{\partial x^2\partial z}
  +\nu \frac{\partial^2 \psi^{[1]}}{\partial z^3}+g\frac{\partial \zeta^{[1]}}{\partial x}-\frac{\sigma }{\rho }\frac{\partial^3 \zeta^{[1]}}{\partial x^3}
=0
\label{g6e1}
\end{eqnarray}
on the surface (\(z=0\)). 

Generally, the first order solution is an integral over wave number \(k\). 

\begin{equation}
  \psi^{[1]} =  \sum_j\int_{-\infty}^\infty dk\; C^{(j)}(k) F^{(j)}(k,z) \exp(i\varphi^{(j)})+c.c.
  \label{xg7a1}
\end{equation}

\begin{equation}
    \zeta^{[1]} = -\sum_j\int_{-\infty}^\infty dk \;C^{(j)}(k)\frac{k}{\Omega^{(j)}} F^{(j)}(k,0)\exp(i\varphi^{(j)})+c.c.\label{xg7a2}
\end{equation}

Where \(\varphi^{(j)}=kx-\Omega^{(j)} t\) and

\begin{dmath}
    F^{(j)}(k,z) = \left(A^{(j)}\cosh k(z+h) -A^{(j)}\cosh\kappa^{(j)} (z+h) +B^{(j)}\sinh \kappa^{(j)} (z+h)
  -\frac{\kappa^{(j)}}{k}B^{(j)}\sinh k(z+h)\right)
\end{dmath}

\begin{eqnarray}
  A^{(j)} &=& 2\kappa^{(j)} k
  \sinh(k h)-\left(\kappa^{(j)2}+k^2\right)\sinh(\kappa^{(j)} h)\nonumber\\
B^{(j)} &=& 2k^2\cosh(k h)-\left(\kappa^{(j)2}+k^2\right)\cosh(\kappa^{(j)} h)\;.
\label{xe26a}
\end{eqnarray}

The angular frequency reads
\begin{eqnarray}
\Omega^{(j)}=i\nu\left(\kappa^{(j)2}-k^2\right)\;.
\label{e26b}
\end{eqnarray}
Modes are labelled by horizontal wave number \(k\) (a continuous parameter) and discrete \(\kappa^{(j)}\) values which are solutions of the following equation \cite{ghahraman2023bifurcation,ref03_hunt1964viscous}
\begin{eqnarray}
&&K\left( Q\sinh K \cosh Q-K\cosh K\sinh
Q \right)(1+sK^2)+p\left[ -4K^2Q\left(K^2+Q^2\right)\right.\nonumber \\&&\left.+Q\left(Q^4+2K^2Q^2+5K^4\right) \cosh
  K\cosh Q -K\left( Q^4+6K^2Q^2+K^4\right) \sinh K\sinh Q \right]=0\;.
\label{e25}
\end{eqnarray}
Here
\begin{eqnarray}
K=kh\;,\label{e23a}\\
Q=\kappa h\;,\label{e23b}\\
p=\frac{\nu^2}{gh^3}\;,
\label{e23c}\\
s=\frac{\sigma}{\rho gh^2}\;.
\label{e23d}
\end{eqnarray}

Summation in (\ref{xg7a1}) and  (\ref{xg7a2}) is understood over the
possible branches (denoted by bracketed upper indices) of the
dispersion relation \(\Omega(k)\). Note that if \(\kappa\) is a solution, so
is \(\kappa^*\) as well at the same \(k\) value, though it may 
yield a different \(\Omega\) branch. The branches of the dispersion relation
at \(k\) are the same as at \(-k\). This allows indexing in the following way:
\begin{eqnarray}
\kappa^{(j)}(-k)&=&\kappa^{(j)*}(k)\nonumber\\
  \Omega^{(j)}(-k)&=&-\Omega^{(j)*}(k)\;.
\end{eqnarray}
Thus we have

\begin{eqnarray}
  \psi^{[1]} = \sum_j\int_{-\infty}^\infty dk \exp(i k x) F^{(j)}(k,z) \times  \left[C^{(j)}(k)+C^{(j)*}(-k)\right]\exp(-i\Omega^{(j)}(k) t)
  \label{xg7b1}\\ 
\end{eqnarray}

\begin{eqnarray}
    \zeta^{[1]} = -\sum_j\int_{-\infty}^\infty dk \exp(i k x)\frac{k}{\Omega^{(j)}} F^{(j)}(k,0)
  \times  \left[C^{(j)}(k)+C^{(j)*}(-k)\right]\exp(-i\Omega^{(j)}(k) t) \label{xg7b2}
\end{eqnarray}

Since the coefficients \(C^{(j)}(k)\) always appear in the combination
\(C^{(j)}(k)+C^{(j)*}(-k)\), henceforth we shall denote this sum with \(C^{(j)}(k)\),
satisfying \(C^{(j)}(-k)=C^{(j)*}(k)\). Note that \(C^{(j)}(k)\) may be expressed in
terms of the spatial Fourier transform  of the initial conditions
\(\zeta^{[1]}(t)\) and \(\dot \zeta^{[1]}(t)\), if we keep just two branches
of the dispersion relation.

\section{Second Order Solutions}
In second order approximation, the main system of equations are

\begin{eqnarray}
  \frac{\partial \triangle \psi^{[2]}}{\partial t}-\nu \triangle \triangle\psi^{[2]}
  =\frac{\partial \psi^{[1]}}{\partial x}\frac{\partial \triangle \psi^{[1]}}{\partial z}
  -\frac{\partial \psi^{[1]}}{\partial z}\frac{\partial \triangle \psi^{[1]}}{\partial x}
\label{g2e2}
\end{eqnarray}
in the bulk, and
\begin{eqnarray}
  \frac{\partial \psi^{[2]}}{\partial x} & = & 0 \label{g3ae2}\\
  \frac{\partial \psi^{[2]}}{\partial z} & = & 0 
\label{g3e2}
\end{eqnarray}
at the bottom (\(z=-h\)). The surface boundary conditions are (at \(z=0\)):
\begin{eqnarray}
\frac{\partial \zeta^{[2]}}{\partial t}-\frac{\partial \psi^{[2]}}{\partial x}=\frac{\partial \psi^{[1]}}{\partial z}\frac{\partial \zeta^{[1]}}{\partial x}+\zeta^{[1]} \frac{\partial^2 \psi^{[1]}}{\partial x\partial z}\;,
\label{g4e2}
\end{eqnarray}

\begin{dmath}
\frac{\partial^2
    \psi^{[2]}}{\partial x^2}-\frac{\partial^2 \psi^{[2]}}{\partial z^2}
=\frac{g}{\nu}\zeta^{[1]}\frac{\partial \zeta^{[1]}}{\partial
  x}
-\frac{\sigma}{\rho\nu}\frac{\partial \zeta^{[1]}}{\partial
  x}
\frac{\partial^2\zeta^{[1]}}{\partial x^2}
\nonumber- \zeta^{[1]} \left(\frac{\partial^3
  \psi^{[1]}}{\partial x^2\partial z}-\frac{\partial^3 \psi^{[1]}}{\partial
  z^3}\right)
-\frac{1}{\nu}\frac{\partial \zeta^{[1]}}{\partial
  x}\int dx\frac{\partial^2 \psi^{[1]}}{\partial t\partial z}
\;, \label{g5e2}
\end{dmath}

and

\begin{dmath}
  -\frac{\partial^2 \psi^{[2]}}{\partial t\partial z}+3\nu \frac{\partial^3
    \psi^{[2]}}{\partial x^2\partial z}
  +\nu \frac{\partial^3 \psi^{[2]}}{\partial z^3}+g\frac{\partial \zeta^{[2]}}{\partial x}-\frac{\sigma }{\rho }\frac{\partial^3 \zeta^{[2]}}{\partial x^3} = \frac{\partial \psi^{[1]}}{\partial x}\frac{\partial^2 \psi^{[1]}}{\partial z^2}-\frac{\partial \psi^{[1]}}{\partial z}\frac{\partial^2 \psi^{[1]}}{\partial x\partial z}
 +\zeta^{[1]}\left[-\frac{\partial^3 \psi^{[1]}}{\partial t\partial
    x^2}+\nu\left(\frac{\partial^4
  \psi^{[1]}}{\partial x^4}-\frac{\partial^4 \psi^{[1]}}{\partial x^2\partial z^2}\right)\right]
+\frac{\partial \zeta^{[1]}}{\partial x}\left[-\frac{\partial^2 \psi^{[1]}}{\partial t\partial
    x}+2\nu\left(\frac{\partial^3
  \psi^{[1]}}{\partial x^3}-\frac{\partial^3 \psi^{[1]}}{\partial x\partial z^2}\right)\right]
+\nu \frac{\partial^2 \zeta^{[1]}}{\partial x^2}\left(\frac{\partial^2
  \psi^{[1]}}{\partial x^2}-\frac{\partial^2 \psi^{[1]}}{\partial z^2}\right)
\;. \label{g6e2}
\end{dmath}

The second order calculation requires evaluation of the right hand side of
Eq.(\ref{g2e2}), which is quadratic in
\(\psi^{[1]}\), therefore, one obtains a double integral. Explicitly, we get for the
rhs. of Eq.(\ref{g2e2})

\begin{eqnarray}
   &&\sum_a\sum_b\int_{-\infty}^\infty dk_1
  \int_{-\infty}^\infty dk_2\;C_1^{(a)}C_2^{(b)}\exp\big\{i (k_1+k_2) x-i\left[\Omega_1^{(a)}+\Omega_2^{(b)}\right] t\big\}\nonumber\\
  &&\times\bigg\{
  \bar \alpha^{(ab)}_{c+}\cosh\left[(k_1+k_2)(z+h)\right]
  +\bar \beta^{(ab)}_{c+}\cosh\left[(\kappa_1^{(a)}+k_2)(z+h)\right]\nonumber\\&&\phantom{\bigg\{}
  +\bar \gamma^{(ab)}_{c+}\cosh\left[(k_1+\kappa_2^{(b)})(z+h)\right]
  +\bar
\delta^{(ab)}_{c+}\cosh\left[(\kappa_1^{(a)}+\kappa_2^{(b)})(z+h)\right]\nonumber\\&& \phantom{\bigg\{} 
  +\bar \alpha^{(ab)}_{c-}\cosh\left[(k_1-k_2)(z+h)\right]
  +\bar \beta^{(ab)}_{c-}\cosh\left[(\kappa_1^{(a)}-k_2)(z+h)\right]\nonumber\\&&\phantom{\bigg\{}
  +\bar \gamma^{(ab)}_{c-}\cosh\left[(k_1-\kappa_2^{(b)})(z+h)\right]
  +\bar \delta^{(ab)}_{c-}\cosh\left[(\kappa_1^{(a)}-\kappa_2^{(b)})(z+h)\right]\nonumber\\&&\phantom{\bigg\{}
  +\bar \alpha^{(ab)}_{s+}\sinh\left[(k_1+k_2)(z+h)\right]
  +\bar \beta^{(ab)}_{s+}\sinh\left[(\kappa_1^{(a)}+k_2)(z+h)\right]\nonumber\\&&\phantom{\bigg\{}
  +\bar \gamma^{(ab)}_{s+}\sinh\left[(k_1+\kappa_2^{(b)})(z+h)\right]
  +\bar \delta^{(ab)}_{s+}\sinh\left[(\kappa_1^{(a)}+\kappa_2^{(b)})(z+h)\right]\nonumber\\&&\phantom{\bigg\{}
  +\bar \alpha^{(ab)}_{s-}\sinh\left[(k_1-k_2)(z+h)\right]
  +\bar \beta^{(ab)}_{s-}\sinh\left[(\kappa_1^{(a)}-k_2)(z+h)\right]\nonumber\\&&\phantom{\bigg\{}
  +\bar \gamma^{(ab)}_{s-}\sinh\left[(k_1-\kappa_2^{(b)})(z+h)\right]
  +\bar \delta^{(ab)}_{s-}\sinh\left[(\kappa_1^{(a)}-\kappa_2^{(b)})(z+h)\right]
  \bigg\}\;.\label{x2e2}
\end{eqnarray}

Note that \(\bar\alpha^{(ab)}_{c\pm}=\bar\alpha^{(ab)}_{s\pm}=0\). Other coefficients
depend on both \(k_1\) and \(\;k_2\) (see Appendix). The solution of
(\ref{g2e2}) can be written in a similar form:

\begin{eqnarray}
  &&\psi^{[2]}=\sum_a\sum_b\int_{-\infty}^\infty dk_1
  \int_{-\infty}^\infty dk_2\;C_1^{(a)}C_2^{(b)}\exp\big\{i (k_1+k_2) x-i\left[\Omega_1^{(a)}+\Omega_2^{(b)}\right] t\big\}\nonumber\\
  &&\times\bigg\{
    \alpha^{(ab)}_{c+}\cosh\left[(k_1+k_2)(z+h)\right]
  +  \beta^{(ab)}_{c+}\cosh\left[(\kappa_1^{(a)}+k_2)(z+h)\right]\nonumber\\&&\phantom{\bigg\{}
  +  \gamma^{(ab)}_{c+}\cosh\left[(k_1+\kappa_2^{(b)})(z+h)\right]
  + 
  \delta^{(ab)}_{c+}\cosh\left[(\kappa_1^{(a)}+\kappa_2^{(b)})(z+h)\right]\nonumber\\&& \phantom{\bigg\{} 
  +  \alpha^{(ab)}_{c-}\cosh\left[(k_1-k_2)(z+h)\right]
  +  \beta^{(ab)}_{c-}\cosh\left[(\kappa_1^{(a)}-k_2)(z+h)\right]\nonumber\\&&\phantom{\bigg\{}
  +  \gamma^{(ab)}_{c-}\cosh\left[(k_1-\kappa_2^{(b)})(z+h)\right]
  +  \delta^{(ab)}_{c-}\cosh\left[(\kappa_1^{(a)}-\kappa_2^{(b)})(z+h)\right]\nonumber\\&&\phantom{\bigg\{}
  +  \alpha^{(ab)}_{s+}\sinh\left[(k_1+k_2)(z+h)\right]
  +  \beta^{(ab)}_{s+}\sinh\left[(\kappa_1^{(a)}+k_2)(z+h)\right]\nonumber\\&&\phantom{\bigg\{}
  +  \gamma^{(ab)}_{s+}\sinh\left[(k_1+\kappa_2^{(b)})(z+h)\right]
  +  \delta^{(ab)}_{s+}\sinh\left[(\kappa_1^{(a)}+\kappa_2^{(b)})(z+h)\right]\nonumber\\&&\phantom{\bigg\{}
  +  \alpha^{(ab)}_{s-}\sinh\left[(k_1-k_2)(z+h)\right]
  +  \beta^{(ab)}_{s-}\sinh\left[(\kappa_1^{(a)}-k_2)(z+h)\right]\nonumber\\&&\phantom{\bigg\{}
  +  \gamma^{(ab)}_{s-}\sinh\left[(k_1-\kappa_2^{(b)})(z+h)\right]
  +  \delta^{(ab)}_{s-}\sinh\left[(\kappa_1^{(a)}-\kappa_2^{(b)})(z+h)\right]
  \bigg\}+\psi^{[2]}_h\;.\label{y2e2}
\end{eqnarray}

Here again \(\alpha^{(ab)}_{c\pm}=\alpha^{(ab)}_{s\pm}=0\). Other coefficients can
be calculated directly from Eq.(\ref{g2e2}) (see Appendix). The homogeneous
part \(\psi^{[2]}_h\) is written as

\begin{eqnarray}
  &&\psi^{[2]}_h=\sum_a\sum_b\int_{-\infty}^\infty dk_1
  \int_{-\infty}^\infty dk_2\;\exp\big\{i (k_1+k_2) x-i\left[\Omega_1^{(a)}+\Omega_2^{(b)}\right] t\big\}\nonumber\\
  &&\times\bigg\{\epsilon_{c}^{(ab)}\cosh\left[(k_1+k_2)(z+h)\right]
  +\epsilon_{s}^{(ab)}\sinh\left[(k_1+k_2)(z+h)\right]\nonumber\\
  &&\;\;
  +\mu_c^{(ab)}\cosh\left[q_{12}^{(ab)}(z+h)\right]+\mu_s^{(ab)}\sinh\left[q_{12}^{(ab)}(z+h)\right]\bigg\}\;,\quad\label{h2e2}
\end{eqnarray}
where
\begin{eqnarray}
  q_{12}^{(ab)2} &=& \kappa_1^{(a)2}+\kappa_2^{(b)2}+2k_1k_2\;.
\end{eqnarray}
Coefficients
\(\epsilon_{c}^{(ab)},\;\epsilon_{s}^{(ab)},\;\mu_c^{(ab)},\;\mu_s^{(ab)}\) are to be
determined from boundary conditions (\ref{g3ae2})-(\ref{g6e2}).

Bottom boundary conditions (\ref{g3ae2}) and (\ref{g3e2}) yield the equations
\begin{eqnarray}
  \epsilon_{c}^{(ab)}+\mu_c^{(ab)}=-\alpha^{(ab)}_{c+}-\beta^{(ab)}_{c+}-\gamma^{(ab)}_{c+}-\delta^{(ab)}_{c+}-\alpha^{(ab)}_{c-}-\beta^{(ab)}_{c-}-\gamma^{(ab)}_{c-}-\delta^{(ab)}_{c-}\label{bbc1}
\end{eqnarray}
and
\begin{dmath}
(k_1+k_2)\epsilon_{s}^{(ab)}+q_{12}^{(ab)}\mu_s^{(ab)}\;= -(k_1+k_2)\alpha^{(ab)}_{s+}-(\kappa_1^{(a)}+k_2)\beta^{(ab)}_{s+}-(k_1+\kappa_2^{(b)})\gamma^{(ab)}_{s+}-(\kappa_1^{(a)}+\kappa_2^{(b)})\delta^{(ab)}_{s+}-(k_1-k_2)\alpha^{(ab)}_{s-}-(\kappa_1^{(a)}-k_2)\beta^{(ab)}_{s-}-(k_1-\kappa_2^{(b)})\gamma^{(ab)}_{s-}-(\kappa_1^{(a)}-\kappa_2^{(b)})\delta^{(ab)}_{s-},\label{bbc2}
\end{dmath}
respectively.

Surface boundary condition (\ref{g4e2}) determines \(\zeta^{[2]}\) as

\begin{eqnarray}
  &&\zeta^{[2]}=\sum_a\sum_b\int_{-\infty}^\infty dk_1
  \int_{-\infty}^\infty dk_2\;C^{(a)}_1C^{(b)}_2\;\exp\big\{i (k_1+k_2) x-i\left[\Omega_1^{(a)}+\Omega_2^{(b)}\right] t\big\}\nonumber\\
  &&\times\bigg\{s^{(ab)}-\frac{k_1+k_2}{\Omega_1^{(a)}+\Omega_2^{(b)}}\bigg(\cosh\left[(k_1+k_2)h\right]\epsilon_{c}^{(ab)}
  +\sinh\left[(k_1+k_2)h\right]\epsilon_{s}^{(ab)}+\cosh\left[q_{12}^{(ab)}h\right]\mu_c^{(ab)}+\sinh\left[q_{12}^{(ab)}h\right]\mu_s^{(ab)}\bigg)\bigg\}\;.\quad\label{sz2e2}
\end{eqnarray}

Here \( s^{(ab)}\) is a function of both \(k_1\) and \(k_2\), its explicite expression is lengthy,
and is given therefore as a supplementary material.

Eqs.(\ref{g5e2}) and (\ref{g6e2}), combined with Eqs.(\ref{bbc1}),
(\ref{bbc2}), (\ref{sz2e2}) provide us with the relations
\begin{eqnarray}
&&
  \bigg\{\left(\kappa_1^{(a)2}+\kappa_2^{(b)2}+k_1^2+k_2^2+4k_1k_2\right)\cosh\left(q_{12}^{(ab)}\right)-2(k_1+k_2)^2\cosh\left(k_1+k_2\right)\bigg\}\mu_c^{ab}\nonumber\\
  &+&\bigg\{\left(\kappa_1^{(a)2}+\kappa_2^{(b)2}+k_1^2+k_2^2+4k_1k_2\right)\sinh\left(q_{12}^{(ab)}\right)-2(k_1+k_2)^2\sinh\left(k_1+k_2\right)\bigg\}\mu_s^{(ab)} \nonumber\\&=&r^{(ab)}\quad\label{sz4e2}
\end{eqnarray}
and

\begin{dmath}
  \left\{\phantom{\frac{\left[k_1^2\right]}{k_1^2}}\mkern -36mu 2q_{12}^{(ab)}\left(k_1+k_2\right)^2\sinh\left(q_{12}^{(ab)}\right)-\left( k_1+k_2 \right) \left( \kappa_1^{(a)2}+\kappa_2^{(b)2}+k_1^2+k_2^2+4k_1k_2\right)\sinh\left(k_1+k_2\right)+\frac{\left(k_1+k_2\right)^2\left(1+s\left(k_1+k_2\right)^2\right)}{p\left[k_1^2+k_2^2-\kappa_1^{(a)2}-\kappa_2^{(b)2} \right]} \left[ \cosh\left(k_1+k_2\right)-\cosh\left(q_{12}^{(ab)}\right)\right]
  \right\}\mu_c^{(ab)}+ \left\{
\phantom {\frac{\left[k_1^2\right]}{k_1^2}}\mkern -36mu 2q_{12}^{(ab)} \left( k_1+k_2 \right)^2 \cosh \left(q_{12}^{(ab)}\right)-q_{12}^{(ab)} \left(\kappa_1^{(a)2}+\kappa_2^{(b)2}+k_1^2+k_2^2+4k_1k_2\right)\cosh\left(k_1+k_2\right)+\frac{\left(k_1+k_2\right)\left(1+s\left(k_1+k_2\right)^2\right)}{p\left[k_1^2+k_2^2-\kappa_1^{(a)2}-\kappa_2^{(b)2}\right]}\left[q_{12}^{(ab)}\sinh\left(k_1+k_2\right)-\left(k_1+k_2\right)\sinh\left(q_{12}^{(ab)}\right)\right]\right\}\mu_s^{(ab)}=t^{(ab)}\;,\quad\label{sz5e2}
\end{dmath}
respectively. The right hand sides \(r^{(ab)}\), \(t^{(ab)}\) no longer contain any unknows, they are
well determined functions of \(k_1\) and \(k_2\). Their explicit expressions
are again given as a supplementary material.

Solving Eqs.(\ref{bbc1}), (\ref{bbc2}), (\ref{sz4e2}), (\ref{sz5e2}) we get finally
\begin{dmath}
  \zeta^{[2]}=\sum_a\sum_b\int_{-\infty}^\infty dk_1
  \int_{-\infty}^\infty dk_2\; C^{(a)}_1C^{(b)}_2\;S^{(ab)}(k_1,k_2)\exp\left\{i (k_1+k_2) x-i\left[\Omega_1^{(a)}+\Omega_2^{(b)}\right] t\right\}\;,\quad\label{sz6e2}
\end{dmath}

where
\(  S^{(ab)}(k_1,k_2) \) stands for the second line of
Eq.(\ref{sz2e2}).

\section{Surface wave equation}

Eqs.(\ref{xg7a2}) and (\ref{sz6e2}) yield the solution for the surface
shape up to second order. We discuss now whether and how one can derive a
single wave equation for the surface shape to this order, without reference to
the underlying bulk Navier-Stokes equations. Provided that we keep only a
single pair of \(\Omega\) values of the dispersion relation, this equation - due to translation symmetry - in
\(k\)-space must have the form
\begin{eqnarray}
  &&\frac{\partial^2\tilde \zeta(k)}{\partial t^2}+
  i\left[\Omega^{(1)}(k)+\Omega^{(2)}(k)\right]\frac{\partial\tilde \zeta(k)}{\partial
    t}-\Omega^{(1)}(k)\Omega^{(2)}(k)\tilde \zeta(k)\nonumber\\
  &&+
  \int_{-\infty}^\infty dk' \;\tilde
  W_1(k',k-k')\tilde\zeta(k')\tilde\zeta(k-k')\nonumber\\
  &&+
  \frac{1}{2}\int_{-\infty}^\infty dk' \;\tilde
  W_2(k',k-k')\left(\tilde\zeta(k')\frac{\partial \tilde\zeta(k-k')}{\partial t}+\frac{\partial \tilde\zeta(k')}{\partial t}\tilde\zeta(k-k')\right)\nonumber\\
  &&+
  \frac{1}{2}\int_{-\infty}^\infty dk' \;\tilde
  W_3(k',k-k')\left(\tilde\zeta(k')\frac{\partial \tilde\zeta(k-k')}{\partial t}-\frac{\partial \tilde\zeta(k')}{\partial t}\tilde\zeta(k-k')\right)\nonumber\\
  &&+
  \int_{-\infty}^\infty dk' \;\tilde
  W_4(k',k-k')\frac{\partial \tilde\zeta(k')}{\partial t}\frac{\partial \tilde\zeta(k-k')}{\partial t}=0
\;,\label{surfeq1}
\end{eqnarray}

where functions  \(\tilde W_1(k_1,k_2)\), \(\tilde W_2(k_1,k_2)\) and \(\tilde
W_{4}(k_1,k_2)\) are symmetric, while \(\tilde
W_3(k_1,k_2)\) is antisymmetric
with respect to the exchange of their variables:
\begin{eqnarray}
  \tilde W_1(k_1,k_2) &=& \tilde W_1(k_2,k_1)\nonumber\\
  \tilde W_2(k_1,k_2) &=& \tilde W_2(k_2,k_1)\nonumber\\
  \tilde W_4(k_1,k_2) &=& \tilde W_4(k_2,k_1)\nonumber\\
  \tilde W_3(k_1,k_2) &=& -\tilde W_3(k_2,k_1)\;.
\end{eqnarray}
Alternatively, one may write
\begin{eqnarray}
  &&\frac{\partial^2\tilde \zeta(k)}{\partial t^2}+
  i\left[\Omega^{(1)}(k)+\Omega^{(2)}(k)\right]\frac{\partial\tilde \zeta(k)}{\partial
    t}-\Omega^{(1)}(k)\Omega^{(2)}(k)\tilde \zeta(k)\nonumber\\
  &&+
  \int_{-\infty}^\infty dk' \;\tilde
  W_{--}(k',k-k')\tilde\zeta(k')\tilde\zeta(k-k')\nonumber\\
  &&+
  2\int_{-\infty}^\infty dk' \;\tilde
  W_{-+}(k',k-k')\tilde\zeta(k')\frac{\partial \tilde\zeta(k-k')}{\partial t}\nonumber\\
  &&+
  \int_{-\infty}^\infty dk' \;\tilde
  W_{++}(k',k-k')\frac{\partial \tilde\zeta(k')}{\partial t}\frac{\partial \tilde\zeta(k-k')}{\partial t}=0
\;,\label{surfeq1a}
\end{eqnarray}
where
\begin{eqnarray}
  \tilde W_{--}(k_1,k_2) &=&\tilde W_1(k_1,k_2)\;,\nonumber\\
  \tilde W_{-+}(k_1,k_2) &=& \frac{1}{2}\left(\tilde W_2(k_1,k_2)+\tilde W_3(k_1,k_2)\right)\;,\nonumber\\
  \tilde W_{+-}(k_1,k_2) &=& \frac{1}{2}\left(\tilde W_2(k_1,k_2)-\tilde W_3(k_1,k_2)\right)\;,\nonumber\\
  \tilde W_{++}(k_1,k_2) &=& \tilde W_4(k_1,k_2)\;.
\end{eqnarray}

We show how the two-variate functions \(\tilde W_j\) may be determined if \(S^{(ab)}\) is known. The idea is to solve (\ref{surfeq1}) perturbatively and
compare the result with Eq.(\ref{sz6e2}). The first order solution of
(\ref{surfeq1}) coincides with the spatial Fourier transform 
of Eq.(\ref{xg7b2}),
\begin{eqnarray}
  \tilde \zeta^{[1]} (k)  =
\sum_aF^{(a)}\;  \exp(-i\Omega^{(a)}(k) t)
  \;,\label{sg7a2}
\end{eqnarray}
where
\begin{eqnarray}
  F^{(a)}  =
-C^{(a)}\frac{k}{\Omega^{(a)}}\left(A^{(a)}\cosh kh -A^{(a)}\cosh\kappa^{(a)} h +B^{(a)}\sinh \kappa^{(a)} h
  -\frac{\kappa^{(a)}}{k}B^{(a)}\sinh k h\right)
  \;.\label{sg7a2b}
\end{eqnarray}
Certainly, the expression of the coefficients is arbitrary, so the specific
form (\ref{sg7a2b}) may be chosen without resticting generality. It is suitable for making contact with the
previous calculations.

The second order
solution satisfies
\begin{eqnarray}
&&\frac{\partial^2\tilde \zeta^{[2]}(k)}{\partial t^2}+
  i\left[\Omega^{(1)}(k)+\Omega^{(2)}(k)\right]\frac{\partial\tilde \zeta^{[2]}(k)}{\partial
    t}-\Omega^{(1)}(k)\Omega^{(2)}(k)\tilde \zeta^{[2]}(k)\nonumber\\
&&+
  \int_{-\infty}^\infty dk' \;\tilde
  W_{1}(k',k-k')\tilde\zeta^{[1]}(k')\tilde\zeta^{[1]}(k-k')\nonumber\\
  &&+
  \frac{1}{2}\int_{-\infty}^\infty dk' \;\tilde
  W_2(k',k-k')\left(\tilde\zeta^{[1]}(k')\frac{\partial \tilde\zeta^{[1]}(k-k')}{\partial t}+\frac{\partial \tilde\zeta^{[1]}(k')}{\partial t}\tilde\zeta^{[1]}(k-k')\right)\nonumber\\
  &&+
  \frac{1}{2}\int_{-\infty}^\infty dk' \;\tilde
  W_3(k',k-k')\left(\tilde\zeta^{[1]}(k')\frac{\partial \tilde\zeta^{[1]}(k-k')}{\partial t}-\frac{\partial \tilde\zeta^{[1]}(k')}{\partial t}\tilde\zeta^{[1]}(k-k')\right)\nonumber\\
  &&+
  \int_{-\infty}^\infty dk' \;\tilde
  W_{4}(k',k-k')\frac{\partial \tilde\zeta^{[1]}(k')}{\partial t}\frac{\partial \tilde\zeta^{[1]}(k-k')}{\partial t}=0
\;.\label{surfeq2a}
\end{eqnarray}

The homogeneous part of the solution may be included in the first order
solution, hence it suffices to construct the particular solution having the
same time dependence as the correction term, i.e.
\begin{dmath}
  \tilde\zeta^{[2]}(k)  = \sum_a\sum_b \int_{ \infty}^\infty dk' \; C^{(a)}(k')C^{(b)}(k-k')\hat S^{(ab)}(k',k-k')\;
  \exp\left(-i\left[\Omega^{(a)}(k')+\Omega^{(b)}(k-k')\right]t\right)  \;,\label{zeta2s1}
\end{dmath}
which is identical with Eq.(\ref{sz6e2}). Upon inserting Eqs.(\ref{sg7a2}) and
(\ref{zeta2s1})
into Eq.(\ref{surfeq2a}) we have
\begin{eqnarray}
  \tilde W_{--}-i\Omega^{(a)}(k')\tilde W_{+-}-i\Omega^{(b)}(k-k')\tilde W_{-+}-\Omega^{(a)}(k')\Omega^{(b)}(k-k')\tilde W_{++}=R^{(ab)}(k',k-k')
    \;,\label{SW}
\end{eqnarray}
where
\begin{eqnarray}
  &&R^{(ab)}(k',k-k') = - \frac{C^{(a)}(k')C^{(b)}(k-k')}{F^{(a)}(k')F^{(b)}(k-k')}\hat
  S^{(ab)}(k',k-k')\nonumber\\
  &&\times\left\{-\left[\Omega^{(a)}(k')+\Omega^{(b)}(k-k')\right]^2+\left[\Omega^{(1)}(k)+\Omega^{(2)}(k)\right]\left[\Omega^{(a)}(k')+\Omega^{(b)}(k-k')\right]-\Omega^{(1)}(k)\Omega^{(2)}(k)\right\}  
    \;.\label{SWa}
\end{eqnarray}

Note that \(R^{(ab)}\) is independent of \(C^{(a)}\) and \(C^{(b)}\), due to
(\ref{sg7a2b}).

Since branch indices \(a\) and \(b\) can be both  either \(1\) or \(2\), the
set of equations  (\ref{SW}) can be solved for any given \(k\) and \(k'\) values:
\begin{eqnarray}
  \tilde W_{--} &=& \frac{\Omega^{(2)}(k')\Omega^{(2)}(k-k')R^{(11)}+\Omega^{(1)}(k')\Omega^{(1)}(k-k')R^{(22)}}{\left[\Omega^{(1)}(k')-\Omega^{(2)}(k')\right]\left[\Omega^{(1)}(k-k')-\Omega^{(2)}(k-k')\right]}\nonumber\\
&&- \frac{\Omega^{(2)}(k')\Omega^{(1)}(k-k')R^{(12)}+\Omega^{(1)}(k')\Omega^{(2)}(k-k')R^{(21)}}{\left[\Omega^{(1)}(k')-\Omega^{(2)}(k')\right]\left[\Omega^{(1)}(k-k')-\Omega^{(2)}(k-k')\right]}\nonumber\\
  \tilde W_{-+} &=& i\frac{\Omega^{(1)}(k')\left(R^{(21)}-R^{(22)}\right)+\Omega^{(2)}(k')\left(R^{(12)}-R^{(11)}\right)}{\left[\Omega^{(1)}(k')-\Omega^{(2)}(k')\right]\left[\Omega^{(1)}(k-k')-\Omega^{(2)}(k-k')\right]}\nonumber\\
  \tilde W_{+-} &=& i\frac{\Omega^{(1)}(k-k')\left(R^{(212)}-R^{(22)}\right)+\Omega^{(2)}(k-k')\left(R^{(21)}-R^{(11)}\right)}{\left[\Omega^{(1)}(k')-\Omega^{(2)}(k')\right]\left[\Omega^{(1)}(k-k')-\Omega^{(2)}(k-k')\right]}\nonumber\\
  \tilde W_{++} &=& \frac{R^{(12)}+R^{(21)}-R^{(11)}-R^{(22)}}{\left[\Omega^{(1)}(k')-\Omega^{(2)}(k')\right]\left[\Omega^{(1)}(k-k')-\Omega^{(2)}(k-k')\right]} 
    \;.\label{Wsol}
\end{eqnarray}
With this, Eq.(\ref{surfeq1a}) is completely defined.

\section{Conclusion}
Second order perturbation theory of the Navier-Stokes equations for free surface flows was presented, wave amplitude considered as the perturbation parameter. The results allowed a systematic derivation of a nonlinear surface wave equation. No assumption for wavelength to layer width ratio was made. Hence, the equation takes into account dispersion fully, while nonlinearity is included in leading order. 

It should be noted that while the concept presented here can be extended to higher orders, one limitation is that we have neglected the presence of the infinitely many overdamped modes and kept only the two least damped ones. Due to nonlinear coupling among the modes, this means that our equation may lead initially, when nonlinearity is significant, to somewhat slower decay than in the reality.  The extent of this should be still estimated.

The results contain rather long expressions, that are difficult to handle. Therefore, discussion of special cases - like weakly damped waves - requires further efforts and is beyond the scope of the present paper.  A further important open question is how our equation can be solved in a numerically effective way.


\section{Acknowledgement}
This research was supported by the Ministry of Culture and Innovation and the National Research and doctoral school of physics at Eötvös Loránd University. A.Gh. greatly acknowledges the support from Tempus Public Foundation for Stipendium Hungaricum Scholarship No. 228712.

\clearpage

\bibliographystyle{unsrt}
\bibliography{main}
\clearpage

\section{Appendix}
Coefficients in Eq.(\ref{x2e2}):
\begin{eqnarray}
 \bar \alpha^{(ab)}_{c+} &=& 0\nonumber\\
  \bar \beta^{(ab)}_{c+} &=&
  \frac{i}{4}\left(k_1-\kappa_1^{(a)}\right)\left(k_1^2-\kappa_1^{(a)2}\right)\left( k_2 A_2^{(b)} B_1^{(a)}+ \kappa_2^{(b)}A_1^{(a)} B_2^{(b)}\right)\nonumber\\
  \bar \gamma^{(ab)}_{c+} &=& \frac{i}{4}\left(k_2-\kappa_2^{(b)}\right)\left(k_2^2-\kappa_2^{(b)2}\right)\left( k_1 A_1^{(a)} B_2^{(b)}+\kappa_1^{(a)} A_2^{(b)} B_1^{(a)}\right)\nonumber\\
  \bar \delta^{(ab)}_{c+} &=& -\frac{i}{4}\left(k_2^2-k_1^2+\kappa_1^{(a)2}-\kappa_2^{(b)2}\right)\left(\kappa_1^{(a)}k_2-\kappa_2^{(b)}k_1\right)\left( A_1^{(a)} B_2^{(b)}+ A_2^{(b)} B_1^{(a)}\right)\nonumber\\
  \bar \alpha^{(ab)}_{c-} &=& 0\nonumber\\
  \bar \beta^{(ab)}_{c-} &=&
  -\frac{i}{4}\left(k_1+\kappa_1^{(a)}\right)\left(k_1^2-\kappa_1^{(a)2}\right)\left( k_2 A_2^{(b)} B_1^{(a)}- \kappa_2^{(b)}A_1^{(a)} B_2^{(b)}\right)
 \nonumber\\
 \bar \gamma^{(ab)}_{c-} &=&
-\frac{i}{4}\left(k_2+\kappa_2^{(b)}\right)\left(k_2^2-\kappa_2^{(b)2}\right)\left( k_1 A_1^{(a)} B_2^{(b)}-\kappa_1^{(a)} A_2^{(b)} B_1^{(a)}\right)
\nonumber\\
\bar \delta^{(ab)}_{c-} &=&
\frac{i}{4}\left(k_2^2-k_1^2+\kappa_1^{(a)2}-\kappa_2^{(b)2}\right)\left(\kappa_1^{(a)}k_2+\kappa_2^{(b)}k_1\right)\left( A_1^{(a)} B_2^{(b)}- A_2^{(b)} B_1^{(a)}\right)
\nonumber\\
\bar \alpha^{(ab)}_{s+} &=& 0\nonumber\\
\bar \beta^{(ab)}_{s+} &=&
  -\frac{i}{4}\left(k_1-\kappa_1^{(a)}\right)\left(k_1^2-\kappa_1^{(a)2}\right)\left( k_2A_1^{(a)} A_2^{(b)} + \kappa_2^{(b)}B_1^{(a)} B_2^{(b)}\right)\nonumber\\
  \bar \gamma^{(ab)}_{s+} &=& -\frac{i}{4}\left(k_2-\kappa_2^{(b)}\right)\left(k_2^2-\kappa_2^{(b)2}\right)\left( k_1 A_1^{(a)} A_2^{(b)}+\kappa_1^{(a)} B_1^{(a)}B_2^{(b)} \right)\nonumber\\
  \bar \delta^{(ab)}_{s+} &=& \frac{i}{4}\left(k_2^2-k_1^2+\kappa_1^{(a)2}-\kappa_2^{(b)2}\right)\left(\kappa_1^{(a)}k_2-\kappa_2^{(b)}k_1\right)\left( A_1^{(a)} A_2^{(b)}+ B_1^{(a)}B_2^{(b)} \right)\nonumber\\
  \bar \alpha^{(ab)}_{s-} &=& 0\nonumber\\
  \bar \beta^{(ab)}_{s-} &=&
  \frac{i}{4}\left(k_1+\kappa_1^{(a)}\right)\left(k_1^2-\kappa_1^{(a)2}\right)\left(
  k_2 A_1^{(a)} A_2^{(b)} - \kappa_2^{(b)}A_1^{(a)} A_2^{(b)}\right)
 \nonumber\\
 \bar \gamma^{(ab)}_{s-} &=&
-\frac{i}{4}\left(k_2+\kappa_2^{(b)}\right)\left(k_2^2-\kappa_2^{(b)2}\right)\left( k_1 A_1^{(a)} A_2^{(b)}-\kappa_1^{(a)} B_1^{(a)} B_2^{(b)}\right)
\nonumber\\
\bar \delta^{(ab)}_{s-} &=&
\frac{i}{4}\left(k_2^2-k_1^2+\kappa_1^{(a)2}-\kappa_2^{(b)2}\right)\left(\kappa_1^{(a)}k_2+\kappa_2^{(b)}k_1\right)\left(
A_1^{(a)} A_2^{(b)}- B_1^{(a)} B_2^{(b)} \right)
\end{eqnarray}
Coefficients in Eq.(\ref{y2e2}):
\begin{eqnarray}
    \alpha^{(ab)}_{c+} &=& 0\nonumber\\
    \beta^{(ab)}_{c+} &=&
  \frac{i}{4\nu}\frac{\left(k_1^2-\kappa_1^{(a)2}\right)\left( k_2 A_2^{(b)} B_1^{(a)}+ \kappa_2^{(b)}A_1^{(a)} B_2^{(b)}\right)}{\left(k_1+2k_2+\kappa_1^{(a)}\right)\left(k_2^2-2k_1k_2+2\kappa_1^{(a)}k_2-\kappa_2^{(b)2}\right)}\nonumber\\
    \gamma^{(ab)}_{c+} &=& \frac{i}{4\nu}\frac{\left(k_2^2-\kappa_2^{(b)2}\right)\left( k_1 A_1^{(a)} B_2^{(b)}+\kappa_1^{(a)} A_2^{(b)} B_1^{(a)}\right)}{\left(k_2+2k_1+\kappa_2^{(b)}\right)\left(k_1^2-2k_1k_2+2\kappa_2^{(b)}k_1-\kappa_1^{(a)2}\right)}\nonumber\\
    \delta^{(ab)}_{c+} &=& \frac{i}{8\nu}\frac{\left(k_2^2-k_1^2+\kappa_1^{(a)2}-\kappa_2^{(b)2}\right)\left(\kappa_1^{(a)}k_2-\kappa_2^{(b)}k_1\right)\left( A_1^{(a)} B_2^{(b)}+ A_2^{(b)} B_1^{(a)}\right)}{\left[\left(k_1+k_2\right)^2-\left(\kappa_1^{(a)}+\kappa_2^{(b)}\right)^2\right]\left(k_1k_2-\kappa_1^{(a)}\kappa_2^{(b)}\right)}\nonumber\\
    \alpha^{(ab)}_{c-} &=& 0\nonumber\\
    \beta^{(ab)}_{c-} &=&
  -\frac{i}{4\nu}\frac{\left(k_1^2-\kappa_1^{(a)2}\right)\left( k_2 A_2^{(b)} B_1^{(a)}- \kappa_2^{(b)}A_1^{(a)} B_2^{(b)}\right)}{\left(k_1+2k_2-\kappa_1^{(a)}\right)\left(k_2^2-2k_1k_2-2\kappa_1^{(a)}k_2-\kappa_2^{(b)2}\right)}
 \nonumber\\
   \gamma^{(ab)}_{c-} &=&
-\frac{i}{4\nu}\frac{\left(k_2^2-\kappa_2^{(b)2}\right)\left( k_1 A_1^{(a)} B_2^{(b)}-\kappa_1^{(a)} A_2^{(b)} B_1^{(a)}\right)}{\left(k_2+2k_1-\kappa_2^{(b)}\right)\left(k_1^2-2k_1k_2-2\kappa_2^{(b)}k_1-\kappa_1^{(a)2}\right)}
\nonumber\\
  \delta^{(ab)}_{c-} &=&
-\frac{i}{8\nu}\frac{\left(k_2^2-k_1^2+\kappa_1^{(a)2}-\kappa_2^{(b)2}\right)\left(\kappa_1^{(a)}k_2+\kappa_2^{(b)}k_1\right)\left( A_1^{(a)} B_2^{(b)}- A_2^{(b)} B_1^{(a)}\right)}{\left[\left(k_1+k_2\right)^2-\left(\kappa_1^{(a)}-\kappa_2^{(b)}\right)^2\right]\left(k_1k_2+\kappa_1^{(a)}\kappa_2^{(b)}\right)}
\nonumber\\
  \alpha^{(ab)}_{s+} &=& 0\nonumber\\
  \beta^{(ab)}_{s+} &=&
  -\frac{i}{4\nu}\frac{\left(k_1^2-\kappa_1^{(a)2}\right)\left( k_2A_1^{(a)} A_2^{(b)} + \kappa_2^{(b)}B_1^{(a)} B_2^{(b)}\right)}{\left(k_1+2k_2+\kappa_1^{(a)}\right)\left(k_2^2-2k_1k_2+2\kappa_1^{(a)}k_2-\kappa_2^{(b)2}\right)}\nonumber\\
    \gamma^{(ab)}_{s+} &=& -\frac{i}{4\nu}\frac{\left(k_2^2-\kappa_2^{(b)2}\right)\left( k_1 A_1^{(a)} A_2^{(b)}+\kappa_1^{(a)} B_1^{(a)}B_2^{(b)} \right)}{\left(k_2+2k_1+\kappa_2^{(b)}\right)\left(k_1^2-2k_1k_2+2\kappa_2^{(b)}k_1-\kappa_1^{(a)2}\right)}\nonumber\\
    \delta^{(ab)}_{s+} &=& -\frac{i}{8\nu}\frac{\left(k_2^2-k_1^2+\kappa_1^{(a)2}-\kappa_2^{(b)2}\right)\left(\kappa_1^{(a)}k_2-\kappa_2^{(b)}k_1\right)\left( A_1^{(a)} A_2^{(b)}+ B_1^{(a)}B_2^{(b)} \right)}{\left[\left(k_1+k_2\right)^2-\left(\kappa_1^{(a)}+\kappa_2^{(b)}\right)^2\right]\left(k_1k_2-\kappa_1^{(a)}\kappa_2^{(b)}\right)}\nonumber\\
    \alpha^{(ab)}_{s-} &=& 0\nonumber\\
    \beta^{(ab)}_{s-} &=&
  \frac{i}{4\nu}\frac{\left(k_1^2-\kappa_1^{(a)2}\right)\left(
  k_2 A_1^{(a)} A_2^{(b)} - \kappa_2^{(b)}A_1^{(a)} A_2^{(b)}\right)}{\left(k_1+2k_2-\kappa_1^{(a)}\right)\left(k_2^2-2k_1k_2-2\kappa_1^{(a)}k_2-\kappa_2^{(b)2}\right)}
 \nonumber\\
   \gamma^{(ab)}_{s-} &=&
-\frac{i}{4\nu}\frac{\left(k_2^2-\kappa_2^{(b)2}\right)\left( k_1 A_1^{(a)} A_2^{(b)}-\kappa_1^{(a)} B_1^{(a)} B_2^{(b)}\right)}{\left(k_2+2k_1-\kappa_2^{(b)}\right)\left(k_1^2-2k_1k_2-2\kappa_2^{(b)}k_1-\kappa_1^{(a)2}\right)}
\nonumber\\
  \delta^{(ab)}_{s-} &=&
-\frac{i}{8\nu}\frac{\left(k_2^2-k_1^2+\kappa_1^{(a)2}-\kappa_2^{(b)2}\right)\left(\kappa_1^{(a)}k_2+\kappa_2^{(b)}k_1\right)\left(
A_1^{(a)} A_2^{(b)}- B_1^{(a)} B_2^{(b)} \right)}{\left[\left(k_1+k_2\right)^2-\left(\kappa_1^{(a)}-\kappa_2^{(b)}\right)^2\right]\left(k_1k_2+\kappa_1^{(a)}\kappa_2^{(b)}\right)}
\end{eqnarray}

\end{document}